\documentclass{elsart}
\usepackage{epsfig}
\usepackage{psfrag}
\newcommand{\beq}{\begin{eqnarray}}
\newcommand{\eeq}{\end{eqnarray}}
\newcommand{\dt}{\displaystyle}
\newcommand{\dfrac}[2]{\frac{\dt #1}{\dt #2}}
\begin{document}

\begin{frontmatter}

\title{Phase transition in fluctuating branched geometry  }

\author[Amsterdam]{Piotr Bialas\thanksref{pperm}},
\author[Bielefeld]{Zdzislaw Burda\thanksref{zperm}}

\address[Amsterdam]{
{Universiteit van Amsterdam, Instituut voor Theoretische
Fysica,}\\ 
{ Valckenierstraat 65, 1018 XE Amsterdam, The
Netherlands}} 
\thanks[pperm]{Permanent address: Institute of Comp. Science,
Jagellonian University,\\ ul. Nawojki 11, 30-072 Krak\'ow, Poland}
\address[Bielefeld]{Fakult\"{a}t f\"{u}r Physik, Universit\"{a}t
Bielefeld,\\ Postfach 10 01 31, Bielefeld 33501, Germany}
\thanks[zperm]{Permanent address: Institute of Physics, 
Jagellonian University, ul. Reymonta 4, 30-059 Krak\'{o}w, Poland}

\begin{abstract}
We study grand--canonical and canonical 
properties of the model of branched polymers proposed in \cite{adfo}.
We show that the model has a fourth order phase
transition and calculate critical exponents.
At the transition the exponent $\gamma$ of the grand-canonical ensemble,
analogous to the string susceptibility exponent of 
surface models, $\gamma \sim 0.3237525...$ is the first known example
of positive $\gamma$ which is not of the form $1/n,\, n=2,3,\ldots$.
We show that a slight modification of the model 
produces a continuos spectrum of $\gamma$'s in the range $(0,1/2]$
and changes the order of the transition. 
\end{abstract}

\end{frontmatter}

\section*{Introduction}
The problem of summing over random geometry appears in 
many areas of modern physics such as string theory, quantum
gravity, membranes and others. The problem is known 
for being hard to study both analytically and numerically.
Therefore one frequently uses simplified models.
In many cases branched polymers capture some essential
futures of more complicated models.
Being simple and solvable they offer us insight into such
issues like value of the string  susceptibility exponent, 
correlation functions, renormalization group, sum over genera 
{\em etc} \cite{adfo,adj,b,abj}.

In the present paper we study a model with a coupling 
to the branching of the polymers \cite{adfo}. 
The model has two regimes. For small values of the coupling
the exponent $\gamma=1/2$ while for larger values it is
negative and depends linearly on the coupling. 
We extend the analysis from the original paper \cite{adfo},
localize the transition and find  that at this point 
$\gamma\sim 0.3237525...$. 

We also study the canonical ensemble
and show that the model exhibits 
a fourth order phase transition.
A slight modification of the model allowed us to obtain a continuos
spectrum of $\gamma$'s between  $0$ and $1/2$. In the canonical ensemble
this generalized model can exhibit a phase transition of 
an arbitrary order equal or higher than 2.

\section*{The model}

\begin{figure}[h]
\vskip5mm
\begin{center}
\psfrag{t1}{$\e^{-\mu}t_1$}
\psfrag{t2}{$\e^{-\mu}t_2$}
\psfrag{t3}{$\e^{-\mu}t_3$}
\epsfig{file=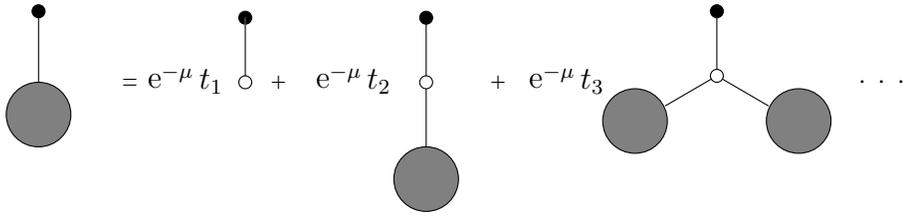,width=12cm}
\end{center}
\caption{\label{Zeqfig} Partition function}
\end{figure}

We consider an ensemble of {\em planar rooted trees} generated
by the recursion relation shown in Fig.\ref{Zeqfig}.
Each bubble in the figure denotes the grand canonical partition 
function $Z$. By iterating the relation one 
produces all possible planar rooted trees. Each tree contributes 
to the partition function the factor $e^{-\mu n}$,
where $n$ is the number of vertices in the tree. 
Each vertex additionally contributes a weight factor depending 
on its degree, {\em ie} the number of branches which meet at vertex. 
In Fig.\ref{Zeqfig} we denoted these weight factors by $t_k$,
where $k$ is degree of vertex. 
In this paper we analyze the
model with $t_k = k^{-\beta}$, where $\beta$ is an external
parameter. The equation from Fig.~\ref{Zeqfig} 
can be written as~: 
\beq
Z(\mu, \beta)  =  
e^{-\mu} \sum_{k=1}^{\infty} \frac{Z(\mu,\beta)^{k-1}}{k^\beta} 
\label{Zdef}
\eeq
which generates following partition function 
(in the following we drop the arguments
of the $Z$ function)
\beq\label{Z}
Z=\sum_{T \in {\mathcal T}} e^{-\mu n(T)} e^{- \beta E(T)}
\eeq
where $\mathcal T$ denotes the ensemble of all trees. 
The function $E(T)$  
plays the role of energy which the tree
contributes to the ensemble and is equal~:
\beq
E(T) = \sum_{v\in T} \ln k(v)
\eeq 
The sum goes over all vertices of the tree $T$, and
$k(v)$ is degree of vertex $v$.
By summing first over trees with fixed number of vertices 
we rewrite (\ref{Z}) as 
\beq
Z = e^{-\phi} = \sum_{n=1}^{\infty} z_n e^{-\mu n} = 
\sum_{n=1}^{\infty} e^{n(f_n - \mu)}
\label{Zseries}
\eeq
The coefficients $z_n$ of the series at 
$e^{-\mu n}$ depend on $\beta$ and correspond to a sum of 
$e^{-\beta E(T)}$ over the ensemble of all trees with $n$ vertices. 
They are canonical partition functions for the ensembles
of trees with $n$ nodes. We have introduced 
convenient variables which we will use later on~: 
free energy density $f_n(\beta)$ for the system with $n$ vertices 
defined from the expansion coefficients~: $z_n = e^{n f_n}$ and 
thermodynamic potential which we call free 
enthalpy $\phi = -\ln Z$. 

For large positive $\beta$ the canonical partition 
function is dominated by branched polymers 
with minimal energy. Such branched polymers 
maximize the number of vertices in the last generation {\em ie} 
vertices with degree $k=1$. They look like short bushes.
In the extreme case such bushes have one branched vertex 
with $n$ branches while all other vertices belong to the last 
generation. The energy of such branched polymers is $\ln n$. 
When $\beta$ becomes smaller the entropy becomes 
important and  more and more elongated structures
appear in the ensemble. At certain $\beta$ the system undergoes 
a fourth order phase transition associated
with the change of the regimes. Finally if one allowed for negative 
$\beta$ then for large negative $\beta$, trees
with large energy would mainly contribute to the
canonical ensemble. The maximal energy $E$ for the
system with $n$ nodes is $n \ln 2$ and comes
from the chain structure.

\section*{Exponent $\gamma$}

The series (\ref{Zseries})
representing the grand canonical 
partition function $Z$ is regular at $e^{-\mu}=0$. 
The behaviour of $z_n$ in the thermodynamic limit 
$n\rightarrow\infty$ can be deduced
from the behaviour of the function $Z$ at 
the radius of convergence $e^{-\mu_0}$, when
$\mu \rightarrow \mu_0^+$,
which corresponds to the first singularity of $Z$
encountered when $\mu$ comes from $\infty$.
At this singularity free enthalpy as a function 
of chemical potential behaves for a given $\beta$ 
as~:
\beq
\phi - \phi_0 \sim (\mu - \mu_0)^{1 - \gamma}
\label{phis}
\eeq
where $\phi_0$, $\mu_0$ and $\gamma$ depend on $\beta$.
The parameter $\mu_0$ is the critical value of the chemical 
potential and $\gamma$ is the universal 
exponent defining a type of singularity which
accounts for the universality class a model belongs to.
Its counterpart in the surface models is called
the string susceptibility exponent. 
As will be seen later $\gamma$
is equal or less than $1/2$ and therefore the function 
on the right hand side goes to $0$ when $\mu$ approaches 
$\mu_0$. The value of free enthalpy at $\mu=\mu_0$ is denoted by 
$\phi_0 = \phi(\mu_0)$. The singularity $(\ref{phis})$ 
corresponds to the following asymptotic behaviour of 
the coefficients $z_n$ in the limit of large $n$~:
\beq
z_n(\beta) \sim 
n^{\gamma-2} e^{ n \mu_0 }
\label{zn}
\eeq
In the thermodynamic limit $n\rightarrow \infty$ the free
energy density is $f_\infty(\beta) = \mu_0(\beta)$.

\begin{figure}[h]
\begin{center}
\psfrag{phi}[lB][lB][1][0]{$\dt e^{-\phi_0}$}
\psfrag{bc}[tl][tl][1][0]{$\beta$}
\epsfig{file=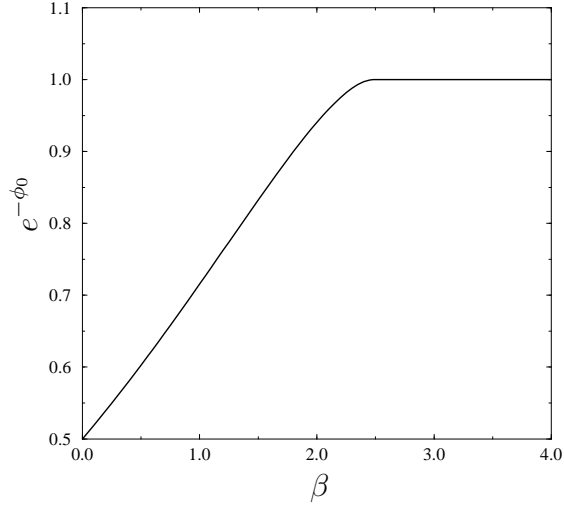,height=7cm}
\end{center}
\caption{\label{phifig} Critical free enthalpy}
\end{figure}

The series $Z(\mu)$ is given by an indirect relation (\ref{Zdef})
which can be rewritten 
\beq
\mu(\phi,\beta) = \ln L_\beta(e^{-\phi}) + 2 \phi
\label{mu}
\eeq
where $L_\beta(z) = \sum_{n=1}^\infty z^n/n^\beta$ is a special
function whose basic properties, frequently used in this paper,
are listed in the appendix. Throughout the paper we use
convention for functions which have $\beta$ 
as second argument that prime denotes derivative 
at fixed $\beta$. 
The function $\phi(\mu)$ is singular at $\mu_0$ 
when either $\mu'(\phi_0) = 0$ or $\mu(\phi_0)$ itself is singular 
at $\phi_0 = \phi(\mu_0)$. 
The first condition reads
\beq
\mu'(\phi_0,\beta) = \frac{\partial \mu(\phi_0,\beta)}{\partial \phi} = 0 \label{phizero}
\eeq
which, by virtue of (\ref{Ldz}), simplifies to~:
\beq
2L_\beta(e^{-\phi_0}) = L_{\beta-1}(e^{-\phi_0})
\label{phicrit}
\eeq
The solution of (\ref{phicrit}) is shown in Fig.~\ref{phifig}.
The function $\phi_0$ approaches its critical value $\phi_C=0$
when $\beta \rightarrow \beta_C$. At this point the
argument of $L_\beta$~: $e^{-\phi_0}$ becomes unity 
and is equal to the radius of convergence of $L_\beta$. 
For $\beta > \beta_C$ the critical value of $\phi$ 
stays at $\phi_C=0$ and corresponds directly to the
singularity of the map $\mu(\phi,\beta)$ .
Inserting $e^{-\phi_0}=1$ to the (\ref{phicrit}) 
we get the critical value of $\beta$~: 
\beq
2\zeta(\beta_C) = \zeta(\beta_C-1)
\eeq
which can be found numerically $\beta_C = 2.4787508... $.

Inserting this back into $\mu$ we get the critical 
value of the chemical potential at the radius of 
convergence of the series (\ref{Zseries}) which also 
corresponds to the free energy density in the 
thermodynamic limit $f_\infty(\beta)$~:
\beq
f_\infty(\beta) = \mu_0 = \mu(\phi_0(\beta),\beta) = \left\{
\renewcommand{\arraystretch}{1.5} 
\begin{array}{lr} 
\ln L_\beta(e^{-\phi_0(\beta)}) + 2\phi_0(\beta),   
&  \quad\beta < \beta_C \\ 
\ln L_\beta(1) = \ln \zeta(\beta), &  \quad\beta \ge \beta_C
\end{array} \right.
\label{f_infty}
\eeq  
We can now compute the  exponent $\gamma$.
We separately consider three cases \\ $\beta < \beta_C$, 
$\beta=\beta_C$ and $\beta > \beta_C$. 
We invert the function $\mu(\phi)$ (\ref{mu})
with respect to $\phi$, for $\phi$ near to $\phi_0$.
For $\beta < \beta_C$ we have~:
\beq
\mu - \mu_0 =  \frac{\mu_0''}{2}(\phi-\phi_0)^2 + \dots 
\eeq
and from there
\beq
(\phi - \phi_0) = 
\bigg(\frac{2}{\mu_0''}\bigg)^{1/2} (\mu - \mu_0)^{1/2} + \dots
\label{ltbc}
\eeq
The second derivative $\mu_0''$ is finite in this range.
For $\beta=\beta_C$ the expansion starts directly from the 
singular term~: 
\beq
\mu - \mu_0 = w(\beta_C) (\phi)^{\beta_C - 1} + \dots 
\label{ebc}
\eeq
because $\mu'=0$ at this point and $\beta_C$ is between $2$ and $3$.
The coefficient $w(\beta) = \Gamma(1-\beta)/\zeta(\beta)$. 
Inverting this with respect to $\phi$ we obtain~:
\beq
\phi = w(\beta_C)^{1/(\beta_C-1)} (\mu - \mu_0)^{1/(\beta_C-1)} + 
\dots
\eeq
Finally, for $\beta>\beta_C$ the expansion of $\mu$ 
around $\mu_0$ starts with a nonvanishing linear part~:
\beq
\mu - \mu_0 =  \mu_0'\phi + \dots + w(\beta) \phi^{\beta - 1} + \dots 
\eeq
The dots between the terms $\phi$ and $\phi^{\beta-1}$
correspond to integer powers $\phi^n$ with $n$ less than $\beta$
(see (\ref{Lx})). Inverted with respect to $\phi$ the equation
yields~:
\beq
\phi = \frac{1}{\mu_0'}(\mu-\mu_0) + \dots - \frac{w(\beta)}{(\mu_0')^\beta}
(\mu-\mu_0)^{\beta-1} + \dots
\label{gtbc}
\eeq
Comparing the singularities with (\ref{phis})  
we obtain~:
\beq
\gamma = \left\{
\renewcommand{\arraystretch}{2}
\begin{array}{lr}
\dt\frac{1}{2}, & \quad\beta < \beta_C \\ 
\dfrac{\beta_C-2}{\beta_C-1} = 0.3237525..., & \quad\beta=\beta_C \\ 
2 - \beta, & \quad\beta > \beta_C 
\end{array}
\right.
\label{gammas}
\eeq

\section*{Canonical ensemble}

Properties of the canonical ensemble in the 
thermodynamic limit are described by the free
energy density $f_\infty(\beta)$. 
It is a continuos function of $\beta$ but
it has singularity at $\beta=\beta_C$.
To extract the dominating singularity of free energy for
$\Delta \beta = (\beta_C - \beta)\rightarrow 0^+$
let us first consider the critical free enthalpy $\phi_0(\beta)$ 
defined by the equation (\ref{phicrit}). 
Using (\ref{Lx}) we can for small $\beta$ and $\phi_0$ 
expand the equation (\ref{phicrit})~:
\beq
- a \Delta \beta + b \phi_0^{\beta_C-2} + ... = 0
\label{ab}
\eeq
up to $o(\Delta \beta)$ and $o(\phi_0^{\beta_C-2})$ .
The expansion coefficients are
$a = 2\zeta'(\beta_C) - \zeta'(\beta_C-1) = 3.4950...$,
$b = -\Gamma(2-\beta_C) = 3.5548...$. 
Solved for $\phi_0$ the last equation gives 
\beq
\phi_0(\beta) =  c \Delta \beta^{3-\eta} + ...
\label{phi0}
\eeq
in the limit $\Delta \beta \rightarrow 0^+$.
The exponent $\eta = 3 - 1/(\beta_C-2) =0.911231...$,
and the coefficient $c=(a/b)^{1/(3-\eta)} = 0.9666...$.
We can now insert $\phi_0$ into (\ref{f_infty}) and compute
singularity of $f_\infty(\beta)$. The first derivative
is~:
\beq 
f_\infty'(\beta) = 
\frac{\partial \mu(\phi_0,\beta)}{\partial \phi} \phi_0' +
\frac{\partial \mu(\phi_0,\beta)}{\partial \beta} = 
\frac{(\partial_\beta L_\beta )(e^{-\phi_0})}{L_\beta(e^{-\phi_0})}
\label{mub}
\eeq
The term containing $\partial \mu/\partial \phi$ 
vanishes identically for $\phi_0$ (\ref{phizero}). 
Now we can expand the last expression in small $\phi_0$, 
and then in small $\Delta \beta$. 
The dominating singularity for $\Delta \beta \rightarrow 0^+$ 
comes from the first nonvanishing term containing $\phi_0$.
This term is linear in $\phi_0$ and introduces the singularity
$\Delta \beta^{3-\eta}$~:
\beq
f_\infty'(\beta) = 
\frac{\zeta'(\beta)}{\zeta(\beta)} + d \Delta \beta^{3-\eta} + ...
\label{defh}
\eeq
The coefficient in front of the singular term~:
\beq
d = c \frac{\zeta'(\beta_C)}{\zeta(\beta_C)}
\Bigg( \frac{\zeta(\beta_C-1)}{\zeta(\beta_C)} - 
\frac{\zeta'(\beta_C-1)}{\zeta'(\beta_C)} \Bigg) = 2.4990...
\nonumber
\eeq
is obtained by expanding (\ref{mub}) and 
using (\ref{Lx}). The first term on the right hand side
of (\ref{defh}) matches the function for 
$\beta > \beta_C$ (\ref{f_infty}) and 
solutions come smoothly one into the other. The 
contribution of the singular term $(\Delta \beta)^{3-\eta}$ 
vanishes at $\Delta \beta=0$ as well as for the second and
the third derivative of the free energy density $f_\infty$. 
For the fourth derivative the singular term blows up at $\beta_C$~:
\beq
\frac{d^4 f_\infty}{d\beta^4}(\beta_C^-) -
\frac{d^4 f_\infty}{d\beta^4}(\beta_C^+)  
= e \Delta \beta^{-\eta} + ...
\label{sing4}
\eeq
where $e = (3-\eta)(2-\eta)(1-\eta) d = 0.5045...$.
This shows that the model has a 
fourth order phase at $\beta_C$ in the canonical sector 
with a singularity given by the exponent $\eta$.

\section*{Non--universality}

The dependence of the exponent $\gamma$ on 
the critical value of the coupling 
suggests a  non--universality. By changing slightly the model
one should be able to change the value of $\beta_C$ without 
spoiling the formulas (\ref{gammas}). In this way one can 
change the value of the exponent $\gamma_C$. We will show below
that indeed a simple modification of the model produces
a whole continuos spectrum of $\gamma$'s.

We modify the model by letting the weight $t_1$, coupled to the
number of points of the last generations, vary in contrast to 
the model defined by (\ref{Zdef}) where it was set to $1$.
Now $t_1$ can take arbitrary value greater than zero. 
The value of $t_1$ affects the number of points in the last
generation. Decreasing/increasing $t_1$ for a given $\beta$ 
favours/suppresses polymers with many points in the last generation. 
By changing $t_1$ we can shift the critical value of $\beta$ 
where the phase transition takes place.
On the other hand this modification does not change the singularity of the 
function on right hand--side of (\ref{Zdef}).

\begin{figure}[h]
\psfrag{t1}{$t_1$}
\psfrag{bc}[lB][lB][1][0]{$\beta_C(t_1)$}
\psfrag{gh2}[bl][bl][1][0]{$\gamma_C=\frac{1}{2}$}
\psfrag{gh1}[bl][bl][1][0]{$\gamma=\frac{1}{2}$}
\psfrag{gb}[bl][bl][1][0]{$\gamma=2-\beta$}
\psfrag{gc}[bl][bl][1][0]{$\gamma_C=\frac{\beta_C-2}{\beta_C-1}$}
\psfrag{beta}{$\beta$}
\begin{center}
\epsfig{file=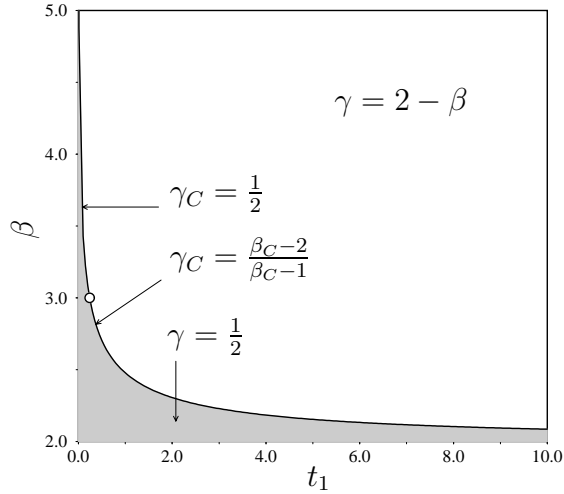,height=7cm}
\end{center}
\caption{\label{bcfig} Phase diagram of the extended model}
\end{figure}

After introducing the modification of the weight $t_1$ the equation
(\ref{mu}) becomes~:
\beq
\mu(\phi,\beta) = \ln \{ L_\beta(e^{-\phi}) +  (t_1-1)\e^{-\phi}\} 
+ 2 \phi
\label{mugen}
\eeq
Repeating the analysis from the previous sections 
we first find the critical line $\phi_0$ as  
a solution of $\mu'=0$~:
\beq
L_\beta(e^{-\phi}) - L_{\beta-1}(e^{-\phi}) = (1-t_1) e^{-\phi}
\eeq
for $\beta$ less than $\beta_C$ and $0$ for $\beta$ 
greater than $\beta_C$. The position 
of the critical point is given by~: 
\beq
2 \zeta(\beta_C) - \zeta(\beta_C-1) = 1-t_1
\eeq
The value $\beta_C$ as a function of $t_1$ is shown in Fig.\ref{bcfig}.
For $t_1\rightarrow +\infty$, the value of $\beta_C$ goes to $2$.
For $t_1\rightarrow 0$, 
$\beta_C$ goes to infinity. 
As first, consider the case $2 < \beta_C < 3$ corresponding to 
$t > t^*_1 = 1+\zeta(2) - 2\zeta(3) = 0.24081\ldots$ . Repeating the
same analysis as for the original model we get 
that $\gamma$ is given exactly by the same formula (\ref{gammas}) 
with $\beta_C$ replaced now by the current value $\beta_C(t_1)$. 
At the critical point we have $\gamma_C = (\beta_C-2)/(\beta_C-1)$ 
which changes contiuosly between $0$, $1/2$ when $\beta_C$ goes 
from $2$ to $3$. When $\beta_C$ exceeds $3$, which is for 
$t_1 < t^*_1$, the situation changes.  
The expansion (\ref{ebc}) starts then not from the singular
term with the power $\beta_C-1$ but from the term 
$\mu''_0 \phi^{2}$ which after inverting gives the singularity 
$\phi \sim (\mu - \mu_0)^{1/2}$ and hence $\gamma=1/2$.

To find the type of the singularity in the canonical ensemble 
we determine the behaviour of the critical free enthalpy 
for $\Delta \beta = (\beta - \beta_C) \rightarrow 0^+$ .
As before first consider the case $2<\beta_C<3$. 
Expanding the equation $\mu'=0$ for small $\Delta \beta$
and small $\phi$ we get exactly the same equation as (\ref{ab})
with exactly the same coefficients but calculated now 
at different $\beta_C$.
The function $\phi_0$ has then the singularity
$\phi_0 \sim (\Delta \beta)^{1/(\beta_C-2)}$ whose power 
depends through $\beta_C$ on the parameter $t_1$ and varies
continuosly.

The situation changes for  $\beta_C > 3$ ($t_1<t^*_1$) because
then additionally regular terms in $\phi_0$ enter the expansion.
They introduce powers of $\phi_0$ smaller than $\phi_0^{\beta_C-2}$~:
\beq
- a\Delta \beta + g \phi_0 +\dots + b \phi_0^{\beta_C-2} + ...  = 0
\label{ab1}
\eeq
where $g = -2\zeta(\beta_C-1) + \zeta(\beta_C-2)$, and $a,b$ are
as in the equation (\ref{ab}). Inverting the last equation 
for $\phi_0$ as a function of $\Delta \beta$ we get~:
\beq
\phi_0 = \frac{a}{g} \Delta\beta + \dots  
- \frac{b}{g} \big(\frac{a}{g}\big)^{(\beta_C-2)}\Delta \beta^{(\beta_C-2)} 
\dots 
\eeq
Inserting the singularity of $\phi_0$ into $f'_\infty$ analogously
as we did in (\ref{defh}) we obtain the singular part of the first 
derivative of the free energy~:
\beq
f'_\infty \sim  
\left\{ \renewcommand{\arraystretch}{1.5}
\begin{array}{lr}
\Delta \beta^{1/(\beta_C-2)}, & \quad t_1 >  t^*_1 \\ 
\Delta \beta^{\beta_C-2}    , & \quad t_1 < t^*_1 
\end{array}
\right.
\eeq
One sees that 
the order of the transition grows arbitraryly and the transitions gets softer 
when $\beta_C$ approaches $2$, corresponding to $t_1 \rightarrow 0$,
or when $\beta_C$,  $t_1 \rightarrow \infty$. The strongest
transition takes place for $t_1 = t^*_1$ for 
which $\beta_C=3$ and the second derivative of the free 
energy diverges logarithmically $\ln (\Delta \beta)$.

\section*{Discussion}

The model described above exhibits a manifold behaviour 
summarized on the phase diagram $(t_1,\beta)$ (Fig.\ref{bcfig}). The 
model has two phases with  bush and tree like branched polymers.
They are separated by a critical line on which the order of 
the transition and the critical exponents vary. 
It has a stable phase for $\beta<\beta_c(t_1)$
where the exponent $\gamma=1/2$ and it is not affected
by a perturbation of couplings. This is a generic branched
polymer phase and has realizations in surface models
and higher simplicial gravity \cite{aj}.  
At the critical line $\beta_C(t_1)$ 
the exponent $\gamma$ is unstable against small 
changes of the model parameters and reveals a continuos spectrum 
in the interval $(0,1/2]$. 

It is known that for $c>1$ exponent $\gamma$ is not universal
which means that $\gamma$ depends not only on $c$ but also on
additional parameters controlling a specific model realization.
For example, for gaussian field on random surface the exponent
$\gamma$ changes as a function on the number of fields $c$ 
and the integration measure parameter \cite{djkp}. 
A classification for a certain subclass of models 
was listed by Durhuus \cite{d} who showed that positive $\gamma$
is of the form $1/n$ $n=2,3,...$ and in particular 
$\gamma=1/3$ is realized by multispin models. 
Our results do not prove that surface models with
arbitrary $\gamma$ in the range $(0,1/2]$ do exist but 
surely make it more plausible.

We believe that models of branched polymers
are worth studying not only in the above mentioned context 
of universality of random surfaces but also as solvable 
models for fluctuating geometry. Our present understanding 
of critical phenomena is mainly based on the spin type models 
on regular fixed geometry. One wonders if this 
intuition can be translated directly 
to fluctuating geometry. In particular, there is a difficulty
with defining correlation functions on fluctuating geometry since
the distance between points fluctuates \cite{b,bs}. 
A study of models as the one discussed here,
especially around the phase transition, can hopefully   
give us a more profound understanding of those problems.

\section*{Acknowledgements}

The authors are grateful to J.~Jurkiewicz,
B.~Petersson and J.~Smit for many valuable discussions.
P.B. thanks the Stichting voor Fundamenteel Onderzoek 
der Materie (FOM) and KBN (grant 2PO3B 196 02  )  
for financial support. Z.B. has 
benefited from the financial support of the
Deutsche Forschungsgemeinschaft under the contract 
Pe 340/3-3.

\section*{Appendix}
We shortly summarize properties of the function defined
by the series~:
\beq
L_\beta(z) = \sum_{n=1}^{\infty} \frac{z^n}{n^\beta}
\label{Ldef}
\eeq
For $z=1$ and $\beta>1$ the function reduces 
to the Riemann Zeta function~:
\beq
L_\beta(1) = \zeta(\beta)
\label{Lzeta}
\eeq
Calculating derivative with respect to the argument $z$
we get within the radius of convergence $|z| < 1$~:
\beq
z\frac{d L_\beta(z)}{dz} = L_{\beta-1}(z) 
\label{Ldz}
\eeq
which also holds for $z=1$ when $\beta>2$. The function
$L_\beta(z)$ has singularity when $z \rightarrow 1$
of the type~:
\beq
\mbox{sing } L_\beta(z) \sim \Gamma(1 -\beta) (1-z)^{\beta-1},\qquad
\beta\ne1,2,\ldots 
\label{Lz1}
\eeq
where $\Gamma$ is the Euler gamma function.

Put together, the last equations give for $x\rightarrow 0$~:
\beq
L_\beta(1-x) = \zeta(\beta) + \dots + (-)^n \zeta(\beta-n) 
\frac{x^n}{n!} +  \Gamma(1-\beta) x^{\beta-1} 
\label{Lx}
\eeq
up to $o(x^{\beta-1})$.
$n$ is the largest integer less than $\beta-1$. 
For integer $\beta$'s the dominating singularity is
of the type $x^{\beta-1} \ln x$.

\end{document}